\documentclass{article}
\usepackage{amsmath}
\usepackage{amsfonts}
\usepackage[latin1]{inputenc}

\newcommand{\di}{\displaystyle}

\newcommand{\R}{\mathbb  R}
\newcommand{\C}{\mathbb  C}
\newcommand{\N}{\mathbb  N}
\newcommand{\1}{\mathbb  I}

\newtheorem{lemma}{Lemma}[section]
\newtheorem{theorem}[lemma]{Theorem}
\newtheorem{proposition}[lemma]{Proposition}

\newtheorem{remark}[lemma]{Remark}

\newtheorem{definition}[lemma]{Definition}

\def\sq{\hbox {\rlap{$\sqcap$}$\sqcup$}}
\def\sq{\hbox {\rlap{$\sqcap$}$\sqcup$}}
\def\noi{\noindent}

\def\beq{\begin{equation}}   \def\eeq{\end{equation}}
\def\bea{\begin{eqnarray}}  \def\eea{\end{eqnarray}}

\def\noi{\noindent}

\newcommand\mysection{\setcounter{equation}{0}\section}
\renewcommand{\theequation}{\thesection.\arabic{equation}}
\newcounter{hran} \renewcommand{\thehran}{\thesection.\arabic{hran}}

\def\bmini{\setcounter{hran}{\value{equation}}
    \refstepcounter{hran}\setcounter{equation}{0}
    \renewcommand{\theequation}{\thehran\alph{equation}}\begin{eqnarray}}

\def\bminiG#1{\setcounter{hran}{\value{equation}}
\refstepcounter{hran}\setcounter{equation}{-1}
\renewcommand{\theequation}{\thehran\alph{equation}}
\refstepcounter{equation}\label{#1}\begin{eqnarray}}

\begin{document}

\title {On  the Herman-Kluk Semiclassical Approximation}
\author{ Didier Robert\\
D\'epartement de Math\'ematiques\\
Laboratoire Jean Leray, CNRS-UMR 6629\\
 Universit\'e de Nantes, 2 rue de la Houssini\`ere, \\
F-44322 NANTES Cedex 03, France\\
\it didier.robert@univ-nantes.fr}
\vskip 1 truecm
\date{}
\maketitle

\begin{abstract}
For a subquadratic symbol   $H$ on $\R^d\times\R^d = T^*(\R^d)$, the quantum   propagator of the time dependent Schr\"odinger equation 
$i\hbar\frac{\partial\psi}{\partial t} = \hat H\psi$   is a Semiclassical Fourier-Integral Operator  when $\hat H=H(x,\hbar D_x)$ ($\hbar$-Weyl quantization of $H$).
Its Schwartz kernel is  describe by a quadratic phase and an amplitude.  At every time $t$, when  $\hbar$  is small,  it  is ``essentially supported"   in a neighborhood of the graph of the classical flow generated by $H$,  with a full  uniform asymptotic expansion  in $\hbar$ for the amplitude.\\
 In this paper our goal is to revisit this well known and fondamental result with emphasis on  the flexibility for the choice of a quadratic complex phase function and on global $L^2$ estimates when $\hbar$ is small  and time $t$ is large. One of the simplest choice of the phase is known in chemical physics as Herman-Kluk formula. Moreover we prove that the semiclassical expansion for the propagator is valid  for $\vert t\vert << \frac{1}{4\delta}\vert\log\hbar\vert$ where $\delta>0$ is a stability parameter for the classical system.

\end{abstract}

\pagestyle{myheadings}

\mysection{Introduction and Results}
Let us consider the time-dependent Schr\"odinger equation 
\beq\label{sch1}
i\hbar\frac{\partial \psi(t)}{\partial t} = \widehat{H}(t)\psi(t),\;\;  
  \psi(t=t_0) = \psi_0, 
\eeq
 where $\psi$  is an initial state, 
  $\hat H(t)$  is 
a quantum Hamiltonian  defined as a continuous family  
 of self-adjoint operators in the Hilbert space $L^2(\R^d)$,  depending on time $t$ and on   the Planck constant  $\hbar > 0$, 
  which plays the role of a small parameter in the   system of units  considered in this paper.  
 $\hat{H}(t)$  is supposed to be  the  $\hbar$-Weyl-quantization   of
 a classical smooth observable  $H(t, X)$, $X=(x, \xi) \in \R^d\times\R^d$ (see \cite{ro1} for  more details concerning semiclassical Weyl quantization). \\
 Our main results concern  subquadratic hamiltonians   $H$;     that means here
  that  $H(t,X)$ is continuous in $t\in\R$, $C^\infty$ smooth in $X\in\R^{2d}$
 and satisfies, for every $\gamma\in\N^{2d}$, $\vert\gamma\vert \geq 2$, 
\beq\label{subq}
\vert\partial_X^\gamma H(t,X\vert \leq C_{T,\gamma},\;\; \forall t,  \vert t-t_0\vert \leq T,\;\; \;\forall X\in\R^{2d}
\eeq
where $\partial_X=\frac{\partial}{\partial X}$ and  $C_{T,\gamma}>0$.\\
Let us introduce  some classes of symbols (``classical observables") defined as follows. Let be $m, n\in\N$.
\begin{definition}
We say that a symbol $s$ is in ${\cal O}_m(n)$ if $s$ is a smooth function on the Euclidian  space $\R^{n}$
such that for every $\gamma\in\N^{n}$, $\vert\gamma\vert \geq m$ we have
\beq
\vert s\vert_{\infty,\gamma}:= \sup_{X\in\R^{n}}\vert\partial_X^\gamma s(X)\vert < +\infty
\eeq
If $s(\varepsilon)$ depends on a parameter $\varepsilon\in P$  we say that  $s(\varepsilon)$ 
is bounded in ${\cal O}_m(n)$ if for every $\gamma$,
we have
$$
\sup_{\varepsilon\in P}\vert s(\varepsilon)\vert_{\infty,\gamma} < +\infty.
$$
\end{definition}

It is well known that the subquadratic assumption entails that equation (\ref{sch1}) is solved by a unique quantum
unitary propagator  in  $L^2(\R^d)$ such that $\psi_t = U(t,t_0)\psi_{0}$, $\forall t\in\R$. 
For the same reason, the classical dynamics is also well defined $\forall t\in\R$. 
$z_t = (q_t, p_t)$ is  the classical path in the phase space $\R^{2d}$ such that $z_{t_0} = z$
 and satisfying
  \beq\label{ham}
 \Bigl\lbrace\begin{array}{l}
 \dot{q_t}  = \partial_p H(t, q_t,p_t) \\ 
   \dot{p_t}  =- \partial H_q(t, q_t,p_t),\;\; q_{t_0} = q,\; p_{t_0} = p
   \end{array}    \eeq
It defines an  Hamiltonian flow : $\phi^{t}(z) = z_t$ ($\phi^{t_0}(z)=z$). 
Let us introduce the stability Jacobi matrix of this Hamiltonian flow : $F(t) = \partial_z\phi^t(z)$.
$F(t)$ is a $2d\times 2d$ symplectic matrix with four $d\times d$ blocks, 
$F(t) = \begin{pmatrix} A_t & B_t\\ C_t&D_t\end{pmatrix}$ , where
 \beq
A_t = \frac{\partial q_t}{\partial q},\; B_t =  \frac{\partial q_t}{\partial p},\; C_t = \frac{\partial p_t}{\partial q},\;
 D_t =  \frac{\partial p_t}{\partial p}
 \eeq
 We also introduce the classical action
 \beq 
 S(t,z)= \int_{t_0}^{t}(p_s\cdot\dot{q_s} - H(s,z_s))ds
  \eeq
  where $u\cdot v$ denote the usual scalar product for $u, v\in\R^d$, 
  and the phase function
  \beq\label{phase1}
 \boxed{\Phi(t,z;x,y) = S(t,z) + p_t\cdot(x-q_t) -p\cdot(y-q) +\frac{i}{2}\bigl(\vert x-q_t\vert^2 + \vert y-q\vert^2\bigr)}
  \eeq
 For applications it is useful to  introduce semi-classical subquadratic symbols. These symbols 
  have an asymptotic expansion in the semiclassical parameter $\hbar >0$, $\di{H^\hbar(t, X)\asymp\sum_{j\geq 0}\hbar^jH_j(t,X)}$
  such that the following conditions are satisfied.
  \bea
  \forall j\geq 0, \; H_j(t,\bullet)\in {\cal O}_{(2-j)_+}(2d) \;\; {\rm and\; are \; bounded\; in}  \;  {\cal O}_{(2-j)_+}(2d) \;{\rm for}\; t\in\R, \\
  \forall N\geq 1,  \hbar^{-N-1}\bigl(H(t, X)-\sum_{0\leq j\leq N}\hbar^jH_j(t,X)\bigr)\; \; {\rm is\;\;bounded\;\;in}\;  {\cal O}_0 \;{\rm for}\;\; t\in\R \;{\rm and }\;\hbar\in]0, 1].
  \eea
  Let us recall the definition of Weyl quantization.  For any symbol $s$ in  ${\cal O}_m(2d)$,and for  any $\psi\in{\cal S}(\R^d)$,
  we have 
  \beq
  Op^w_\hbar[s]\psi(x) = (2\pi\hbar)^{-d}\int\!\!\!\int_{\R^{2d}}{\rm e}^{\frac{i}{\hbar}(x-y)\cdot\xi}s\bigl(\frac{x+y}{2}, \xi\Bigr)\psi(y)dyd\xi.
  \eeq
  We shall also use the notation $\hat s =Op^w_\hbar[s]$.

  The Herman-Kluk formula is included in the following asymptotic result which will be discussed in details in this paper.
  This formula was discovered by several authors in the chemical-physics litterature in the eighties.  We refer to the introductions of 
  \cite{ka2} and \cite{rosw} for  interesting historical expositions.  It is rather surprising that until the recent  paper \cite{rosw} there was no explicite connexion
  in the mathematical litterature between the Herman-Kluk formula and Fourier-Integral Operators with complex phases.
\begin{theorem}\label{Thm1}
Let be $H^\hbar(t)$ a time dependent semiclassical subquadratic Hamiltonian   and $K^\hbar(t;x,y)$  be the Schwartz kernel of 
its propagator $U^\hbar(t,t_0)$.
Then there exists a semi-classical symbol of order 0, $\di{a^\hbar(t;z) = \sum_{0\leq j<+\infty}a_j(t;z)\hbar^j}$ where $a_j$
is continuous in $t$, 
\beq\label{HK1}
K^\hbar(t;x,y) \asymp \int_{\R^{2d}}{\rm e}^{\frac{i}{\hbar}\Phi(t,z;x,y)}a(\hbar;t;z)dz
\eeq
in the $L^2$ uniform norm. More precisely, if we denote 
\beq\label{HKapp}
K^{(\hbar,N)}(t;x,y) = (2\pi\hbar)^{-3d/2}\int_{\R^{2d}}{\rm e}^{\frac{i}{\hbar}\Phi(t,z;x,y)}\bigl(\sum_{0\leq j\leq N}a_j(t;z)\hbar^j\bigr)dz
\eeq
and  $U^{(\hbar, N)}(t,t_0)$ the operator, in $L^2(\R^d)$,  with the Schwartz kernel $K^{(\hbar,N)}(t;x,y)$,  then,
for every $T>0$ and every $N\geq 1$, there exists $C(T,N)>0$ such that for the $L^2$ operator norm we have 
\beq
\Vert U\hbar(t,t_0)- U^{(\hbar,N)}(t,t_0)\Vert \leq C(T,N)\hbar^{N+1},\;\; \forall t, \vert t-t_0\vert\leq T,\; \hbar\in]0, 1].
\eeq
The leading term is
\beq\label{HKT}
a_0(t;z) = {\rm det}^{1/2}(A_t+D_t +i(B_t-C_t))\exp\Bigl(-i\int_{t_0}^tH_1(z_s)ds\Bigr)
\eeq
where the square root is defined by continuity starting from $t=t_0$ ($a_0(t_0;z)=2^{d/2}$).\\
Moreover, the amplitudes $a_j$ are smooth functions defined by transport equations (see the proof below)  and, for every $T>0$ they  
are bounded in ${\cal O}_0$ for $\vert t\vert \leq T$. 
\end{theorem}
In \cite{rosw} the authors give a rigorous proof of this result with an additionnal hypothesis : they
assume that $H(x,\xi)$ is a polynomial in $\xi$. Here we consider more general subquadratic symbols. In particular our result applies to 
 relativistic Hamiltonians like
$\sqrt{1+\vert\xi\vert^2}+V(x)$. Using a global diagonalization  (see \cite{ro2}, section 3), the result can be extended to Dirac systems. \\
 Similar results are true with more general quadratic phases and for systems with diagonalisable leading  symbols (see
 \cite{bi, ro2}). Let us define the quadratic phase
\beq\label{phase2}
\Phi^{(\Theta_t,\Gamma)}(t,z;x,y) = S(t,z) + p_t\cdot(x-q_t) -p\cdot(y-q) +\frac{1}{2}
\bigl(\Theta_t(x-q_t)\cdot(x-q_t) - \overline{\Gamma}(y -q).(y-q)\bigr)
\eeq
where $\Gamma, \Theta_t$ are complex symmetrix matrices with a definite-positive imaginary part, $\Theta_t$ is $C^1$ in $t$. 
$\Gamma$ is constant, $\Theta_t$  may depend smoothly on  $t$ and $z$ such that the following condition is satisfied :
\beq\label{estmat1}
\exists c_T>0,\; \Im\Theta_t v.v \geq \frac{1}{c_T}\vert v\vert^2,\;\; \forall t, \; \vert t\vert\leq T,\;\; \forall z\in\R^{2d}
\eeq
\beq\label{estmat2}
\forall\gamma,  \vert\gamma\vert\geq 1, \exists C_{T,\gamma},\; \Vert\partial_z^\gamma\Theta_t\Vert \leq C_{T,\gamma},\; \forall z\in\R^{2d}, 
\forall \vert t\vert\leq T.
\eeq
So we have
\begin{theorem}\label{PVVK}
Under the assumptions of {\rm Theorem} \ref{Thm1} and  (\ref{estmat1}), (\ref{estmat2}), we have 
\beq\label{VVK}
K(t;x,y) \asymp (2\pi\hbar)^{-3d/2}\int_{\R^{2d}}{\rm e}^{\frac{i}{\hbar}\Phi^{(\Theta_t,\Gamma)}(t,z;x,y)}f(\hbar;t;z)dz
\eeq
where  $\di{f(\hbar;t;z) =\sum_{0\leq j<+\infty}f_j(t;z)\hbar^j}$ with the same meaning as in {\rm Theorem \ref{Thm1}}.\\
In particular 
\beq
f_0(t,z) = 2^{d/2}{\rm det}^{1/2}[M(\Theta_t,\Gamma)\big]
\eeq
where 
$$
M(\Theta_t,\Gamma) = i\bigl(C+D\overline{\Gamma}-\Theta(A+B\overline{\Gamma}\bigr).
$$
\end{theorem}
There exist  several methods to prove this theorem. In \cite{rosw} the authors  prove it as a consequence of a symbolic calculus for FIO with complex quadratic  phases. In \cite{biro} the authors proved a weaker result for $\Gamma=i\1$ and $\Theta_t=\Gamma_t$ is determined by the propagation
of Gaussian coherent states: $ \Gamma_t = (C+D\Gamma)(A+B\Gamma)^{-1}$ (see section 2 of this paper). 
Laptev-Sigal in \cite{lasi} have also considered a similar formula  for the propagator  (see section 5 of this paper)
 but assume that the   initial data  has a compact support in momenta.
Kay \cite{ka2}  explains how to  compute all the semiclassical corrections $a_j$ but did not give estimates on the error term,
so its expansion is not rigorously established. 
Here we choose another approach,  may be more explicit and 
simpler. We shall prove the general theorem \ref{PVVK} as a consequence of the particular case  of Theorem \ref{Thm1} by using
 a real deformation of the phase $\Phi^{(\Theta_t,\Gamma)}$ on the simpler one  $\Phi^{(i\1,i\1)}$. Moreover we give a direct proof of Theorem \ref{Thm1}, proving the necessary properties for Fourier integrals with complex quadratic phases.
  This way we can get easily explicit  estimates for the error terms  for large times.

 Let us assume that conditions  on $\hat H(t)$ are satisfied for $T=+\infty$. Moreover assume that there exists a positive real function
 $\mu(T)\geq 1$, $T>0$,  such that the classical flow $\phi^t$ satisfies, for every multiindex $\gamma$, 
 $\vert\gamma\vert\geq 1$, we have for some $C_\gamma>0$, 
 \beq\label{flowgl}
 \vert\partial_z^\gamma\phi^{t,t^\prime}(z)\vert \leq C_\gamma\mu(T)^{\vert\gamma\vert},\;\;  {\rm for }\;\; \vert t\vert +\vert t^\prime\vert \leq T,\;\;\forall z\in\R^{2d}
 \eeq
 We have discussed in \cite{biro} the condition (\ref{flowgl}). In particular this condition is fulfilled  with $\mu(T) = {\rm e}^{\delta T}$
 for $\di{\delta = \sup_{X\in\R^{2d}, t\in\R}\Vert J\partial_{X,X}^2H(t,X)}\Vert$.
\begin{theorem}\label{eht}
Choosing the phase as in theorem \ref{Thm1}, for $j\geq 0$ the amplitudes $a_j(t,z)$ satisfy the following estimates,
for every multiindex $\gamma$ there exist a constant $C_{j,\gamma}$ such that
\beq
\vert\partial_z^\gamma a_j(t,z)\vert \leq C_{j\gamma}\vert{\rm det}^{1/2}M_t\vert\mu(t)^{4j+\vert\gamma\vert},\;\; \forall t\in\R,\; \forall z\in \R^{2d}.
\eeq
Hence we have the  following Ehrenfest type estimate. For every $N\geq 1$ and every $\varepsilon >0$ there exists
 $C_{N,\varepsilon}$
such that we have 
\beq
\Vert U(t,t_0)- U^{(N)}(t,t_0)\Vert \leq C_{N,\varepsilon}\hbar^{\varepsilon(N+1)},\;\; \forall t, \vert t\vert \geq \frac{1-\varepsilon}{4\delta},\;\forall
\hbar\in]0, 1].
\eeq
\end{theorem}
In previous works an Ehrenfest time  $T_E =  c\log\hbar^{-1}$, $c>0$,  was estimated for propagation of Gaussians in \cite{coro1}
and propagation of observables in \cite{boro}.  For Gaussians we got $c=\frac{1}{6\delta}$, for observables $c=\frac{1}{2\delta}$.
In \cite{rosw}  the authors gave an Ehrenfest time without  explicit estimate on  $c$.

\section{Gaussians Coherent States and Quadratic Hamiltonians}
The phase functions $\Phi^{(\Theta, \Gamma)}$ in (\ref{phase1}) and (\ref{phase2})  are  closely related with Gaussian coherent states. This can be seen by  proving  a particular case of Theorem.\ref{Thm1}  
for   quadratic  time-dependent Hamiltonians: 
      $$
   H_t(q,p) = \frac{1}{2}\left(G_tq\cdot q + 2L_tq\cdot p + K_tp\cdot p\right)
   $$ 
   where $q, p \in\R^d$,  $K_t, L_t, G_t$ are real,  $d\times d$  matrices,
   continuous in time $t\in\R$, $G_t, K_t$ are symmetric. The classical motion in the phase space
    is given by the linear differential equation
    \beq\label{classicevol}
    \left(\begin{array}{c}\dot q\\ \dot p\end{array}\right) 
     = J.\left(\begin{array}{cc}G_t & L_t^T\\
     L_t & K_t \end{array}\right)\left(\begin{array}{c}q \\p\end{array}\right),\;\;\;J = \begin{pmatrix} 0&\1\\-\1&0\end{pmatrix}
     \eeq
     where $L^T$ is the transposed matrix of $L$, $J$  defines the symplectic form $\sigma(X,X^\prime) := JX\cdot X^\prime$,
      $X=(x,\xi)$, $X^\prime=(x^\prime,\xi^\prime)$. \\
     This equation defines a linear symplectic transformation, $F_{t}$, 
      such that  $F_{0} = \1$ (we take here $t_0=0$).    It can be represented        
       as a     $2d\times 2d$ matrix  
  which can be written as four $d\times d$ blocks~:
 \beq 
F_{t}= \left(\begin{array}{cc}A_{t} & B_{t }\\
C_{t}&  D_{t} \end{array}\right).
\eeq
The quantum evolution for the Hamiltonian $\hat H(t)$ is denoted by $U(t)$ ($U(0)=\1$).
We can  compute the matrix elements of $U(t)$
on the coherent states basis  $\varphi_z$. This has been done  in Littlejohn \cite{li} (p.249, (6.36)), 
 Bargmann \cite{ba}, Fedosov \cite{fed}, \cite{coro2}.  We follow  here the presentation given in \cite{coro2}.
 Let us introduce  some notations which will be used later.
$g$ denotes the Gaussian function: $g(x) = \pi^{-d/4}{\rm e}^{-\vert x\vert^2/2}$  and  $\Lambda_\hbar$ is the 
 dilation operator $\Lambda_\hbar\psi(x) = \hbar^{-d/4}\psi(\hbar^{-1/2}x)$. 
 So $\varphi_0 =\Lambda_\hbar g$, and the general Gaussian coherent states are defined as follows.
\beq
   \varphi_{z}^{(\Gamma)} = \hat T(z)\varphi^{(\Gamma)},
   \eeq
   where  $\hat T(z)$ is the Weyl translation  operator, $z = (q, p)$, 
  \beq
\hat T(z) = \exp\left(\frac{i}{\hbar}(p\cdot x - q\cdot\hbar D_x) \right) 
\eeq
where $D_x = -i\frac{\partial}{\partial x}$ and $z = (q,p)\in\R^d\times\R^d$.
$\varphi^{(\Gamma)}$ is the  Gaussian state:
  \beq
  \varphi^{(\Gamma)}(x) = (\pi\hbar)^{-d/4}a_\Gamma\exp\left(\frac{i}{2\hbar}\Gamma x.x\right)
  \eeq
    where 
$\Gamma$ is a complex symmetric matrix such that $\Im\Gamma$ is definite-positive, $a_\Gamma$  is a normalization constant.
($a_\Gamma = {\rm det}^{1/4}\Im\Gamma$).\\
It is convenient to introduce here  the Siegel space  $\Sigma_+(d)$ of  $d\times d$  complex matrices $\Gamma$ 
such that $\Im\Gamma$ is definite-positive. (see in \cite{fo} properties of $\Sigma_+(d)$).

Let us define the  Fourier-Bargmann transform ${\cal F}^{(\Gamma)}_{\cal B}$ as follows, $\psi\in L^2(\R^{d})$, 
\beq
 {\cal F}^{(\Gamma)}_{\cal B}[\psi](z) = (2\pi\hbar)^{-d/2}\langle\psi, \varphi_z^{(\Gamma)}\rangle.
  \eeq 
  $z\in\R^{2d}$, $\varphi^{(\Gamma)}_z$ is the following coherent state living at $z$, 
  $z=(q,p)\in\R^d\times\R^d$, $x\in\R^d$, 
  \beq
  \varphi^{(\Gamma)}_z(x) = (\pi\hbar)^{-d/4}a_\Gamma
  \exp\Bigl(\frac{i}{\hbar}\bigl(p\cdot x - \frac{p\cdot q}{2}\bigr) +\frac{i\Gamma(x-q)\cdot(x-q)}{2\hbar}\Bigr)
   \eeq
 ${\cal F}^{(\Gamma)}_{\cal B}$  is an isometry from  
 $L^2(\R^d)$ into $L^2(\R^{2d})$ (with the Lebesgue measures). If $\Gamma=i\1$ we denote ${\cal F}_{\cal B}={\cal F}^{i\1}_{\cal B}$; 
  its range consists of $F\in L^2(\R^{2d})$ such that
   $\exp\left(\frac{p^2}{2} +i\frac{q\cdot p}{2}\right)F(q,p)$ is holomorphic in $\C^d$ in the variable $q-ip$.  In other words, 
    \beq
    {\mathcal F}_{\mathcal B}\psi(z) = 
     E_\psi(q-ip)\exp\left(-\frac{p^2}{2} -i\frac{q\cdot p}{2}\right)
    \eeq     where $E_\psi$ is entire in $\C^d$ (see \cite{ma}).
Moreover  we have the inversion formula
\beq\label{inv}
 \psi(x)  = \int_{\R^{2d}}{\mathcal F}^{(\Gamma)}_{\mathcal B}[\psi](z)\varphi^{(\Gamma)}_z(x)dz,\;\; {\rm in\; the}\; L^2-{\rm sense}.
\eeq
These properties are well known (see \cite{ma, biro}).
Sometimes we shall use the shorter notation 
$\tilde\psi^{\Gamma} =  {\mathcal F}_{\cal B}^{(\Gamma)}\psi$ and $\tilde\psi^{\Gamma} = \tilde\psi$. 

Let us  denote by $\hat R[F_t]$ the quantum propagator  for the Hamiltonian $H(t)$  (this is the metaplectic representation of $F_t$)
and $K^{(F_t)}$ its Schwartz kernel. 
We know that $\Lambda_\hbar\hat R[F_t]g$ is the following Gaussian state \cite{coro2, fo}, 
 \beq
 \Lambda_\hbar\hat R[F_t]g(x) = (\pi\hbar)^{-d/4}a_\Gamma(t)\exp\left(\frac{i}{2\hbar}\Gamma_t x.x\right)
  \eeq
  where 
$a_\Gamma(t) = [{\rm det}(A_t +\Gamma B_t)]^{-1/2}a_\Gamma$,
 the complex square root is computed by continuity \footnote{ this definition of ${\rm det}^{1/2}$ is different that the ${\rm det}^{1/2}$ function  on $\Sigma_+(d)$,
 this is explained in \cite{coro2} to compute Maslov index }
  from $t = t_0=0$, and 
 \beq
 \Gamma_t =  (C_t + \Gamma D_t)(A_t + \Gamma B_t)^{-1},\;\; \Gamma_{t_0} = \Gamma.
 \eeq
 \begin{proposition}
 We have the following exact formula
 \beq\label{kquad}
 K^{(F_t)}(x,y) = 2^{d/2}(2\pi\hbar)^{-3d/2}{\rm det}^{1/2}\Bigl(\frac{M(\Theta_t,\Gamma)}{i}\Bigr)\int_{\R^{2d}}
 {\rm e}^{\Phi^{(\Theta,\Gamma)}(t,z;x,y)}dz
 \eeq
 where $\Gamma, \Theta_t \in\Sigma_+(d)$,
 $\Theta_t$ is $C^1$ in $t$;  $M(\Theta_t,\Gamma) =  C+D\bar\Gamma - \Theta_t(A+B\bar\Gamma)$ and 
 $$
\Phi^{(\Theta_t,\Gamma)}(t,z;x,y) = \frac{1}{2}(q_t\cdot p_t -q\cdot p) + p_t\cdot(x-q_t) -p\cdot(y-q) +\frac{1}{2}
\bigl(\Theta_t(x-q_t)\cdot(x-q_t) - \overline{\Gamma}(y -q).(y-q)\bigr)
$$
 Let us remark that here   the action is  $S(t,z) = \frac{1}{2}(q_t\cdot p_t -q\cdot p)$.
 \end{proposition}
 First of all let us remark that the integral (\ref{kquad}) is an oscillating integral and is defined, as usual, by integrations by parts.
 We shall give two proofs of this formula.\\
 {\bf Proof I}. We start with any $\Gamma_0$ in the Siegel space $\Sigma_+(d)$. Using  the formula
 $$
 \psi(x) = (2\pi\hbar)^{-d}\int_{\R^{2d}}\langle\psi,\varphi^{\Gamma_0}_z\rangle\varphi^{\Gamma_0}_zdz
 $$
we get the formula
\beq
 K^{(F_t)}(x,y) = (2\pi\hbar)^{-d}\int_{\R^{2d}}\overline{\varphi^{(\Gamma_0}_z(y)}\varphi^{(\Gamma_t)}_{z_t}(x)dz
\eeq
So, we get
\beq
 K^{(F_t)}(x,y)  = (2\pi\hbar)^{-3d/2}k_0(t)\int_{\R^{2d}}{\rm e}^{\frac{i}{\hbar}\Phi^{(\Gamma_t,\Gamma_0)}(t,z;x,y)}dz,
\eeq
where
$$
k_0(t)= 2^{d/2}\frac{{\rm det}^{1/2}(\Im\Gamma_0)}{{\rm det}^{1/2}(A+B\Gamma_0)}
$$
Now we shall   transform the phase $\Phi^{(\Gamma_t,\Gamma_0)}$  into  the phase   $\Phi^{(\Theta,\Gamma_0)}$.\\
Let us introduce $\Theta(s) =  s\Theta +(1-s)\Gamma_t$, $0\leq s\leq1$.  We have $\Theta(s)\in\Sigma_+(d)$.
 We want to find $k(t,s)$ such that
$k(t,0)=k_0(t)$ and 
\beq
\frac{\partial}{\partial s}\Bigl(k(t,s)\int_{\R^{2d}}{\rm e}^{\frac{i}{\hbar}\Phi^{(\Theta_t,\Gamma_0)}(t,z;x,y)}dz\Bigr) = 0,\;\;\forall s\in[0, 1].
\eeq
We have
$$
\frac{\partial}{\partial s}{\rm e}^{\frac{i}{\hbar}\Phi^{(\Theta_t,\Gamma_0)}} =
\frac{i}{2\hbar}(\Theta_t-\Gamma_t)(x-q_t)\cdot(x-q_t){\rm e}^{\frac{i}{\hbar}\Phi^{(\Theta_t, \Gamma_0)}}.
$$
The main trick used here and later in this paper, and also  in all  the previous papers on this subject (\cite{lasi, ka2, rosw}),   is to integrate by parts to convert  each  factor $(x-q_t)$ into $\hbar$, using the following equality 
\beq
(\partial_q +\bar\Gamma\partial_p)\Phi^{\Theta,\Gamma} = 
\bigl(C^\tau +\bar\Gamma D^\tau - (A^\tau +\bar\Gamma B^\tau)\Theta\bigr)(x-q_t) 
\eeq
where $A^\tau$ denotes the transposed matrix of $A$. 
Let us  introduce the matrix
$$
M =  M(\Theta,\Gamma) =  C+D\bar\Gamma - \Theta(A+B\bar\Gamma)
$$
So we have
\beq
M^\tau(x-q_t){\rm e}^{\frac{i}{\hbar}\Phi^{(\Theta,\Gamma)}} = 
\frac{\hbar}{i}\bigl(\partial_q +\bar\Gamma\partial_p\bigr){\rm e}^{\frac{\hbar}{i}\Phi^{\Theta,\Gamma}}.
\eeq
Let us remark that $M$ is invertible. This is a consequence of  the following Lemma (see \cite{cofe}, \cite{fo} or \cite{ro2}, appendix A,  for proofs).
\begin{lemma}\label{nond}
For every linear symplectic  map in $F: T^*(\R^d)\rightarrow T^*(\R^d)$,  $F=\begin{pmatrix} A & B\\ C& D\end{pmatrix}$ and every $\Gamma\in\Sigma_+(d)$, 
 $(A+B\Gamma)$, $(C+D\Gamma)$  are non invertible in $\C^d$  and   $(C+D\Gamma)(A+B\Gamma)^{-1}\in\Sigma_+(d)$. 
\end{lemma}
So we have
$$
\bar M = C+D\Gamma -\bar\Theta(A+B\Gamma) = \bigl((C+D\Gamma)(A+B\Gamma)^{-1} -\bar\Theta\bigr)(A+B\Gamma)^{-1}
$$
But  $(C+D\Gamma)(A+B\Gamma)^{-1} -\bar\Theta)\in\Sigma_+(d)$ so is invertible.\\
Denote $M(t,s) = M(\Theta_s,\Gamma_t)$.  Let us recall the Liouville formula
\beq
\partial_s{\rm det}\bigl(M(t,s)\bigr) = {\rm det}\bigl(M(t,s)\bigr){\rm Tr}\Bigl(\partial_sM(t,s)M(t,s)^{-1}\Bigr).
\eeq
So, integrating by parts in $(q,p)$ we get 
\beq
k(t,s) = k(t,0)\frac{{\rm det}^{1/2}M(t,s)}{{\rm det}^{1/2}M(t,0)}
\eeq
Now we have to compute  $\frac{k(t,0)}{{\rm det}^{1/2}M(t,0)}$.  A simple computation gives
$M(t,0) = (D-\Gamma_tB)(\bar\Gamma_0-\Gamma_0)$.  The proof of (\ref{kquad}) follows from 
the formula
\beq\label{comp}
{\rm det}(D-\Gamma_tB) = {\rm det}(A+B\Gamma_0)^{-1}.
\eeq
This  equality follows from the symplecticity of $F$ ($D^\tau B=B^\tau D$). We have
$B^\tau\Gamma_t B- D^\tau B = -(A+B\Gamma_0)^{-1}B$.  So we get (\ref{comp})  if ${\rm det} B\neq 0$.
The general case follows by a density argument.\\
Let us remark  that  can exchange the  role of $\Theta$ and $\Gamma$
by considering  the adjoint $U(t)^*$ of $U(t)$. \sq

\noi
{\bf Proof II}\\
We solve  directly the Schr\"odinger equation
\beq
\bigl(i\hbar\frac{\partial}{\partial t} -\hat H(t)\bigr)\psi(t,x) = 0
\eeq
for any initial data $\psi(x):=\psi(0,x)$, $\psi\in{\cal S}(\R^d)$
using the ansatz
\beq\label{ansatz}
\psi(t,x) = (2\pi\hbar)^{-3d/2} k(t)\int_{\R^{2d}\times\R^d}{\rm e}^{i\Phi^{(\Theta,\Gamma)}(t,z;x,y)}\psi(y)dzdy
\eeq
We have to compute $k(t)$ such that $k(0) = 2^{d/2}$. Let us remark that if we integrate first in $y$ then the integral 
 (\ref{ansatz}) in $z$
converges because  the Fourier-Bargmann transform of $\psi$,  ${\cal F}_{\cal B}\psi$,   is in the Schwartz space ${\cal S}(\R^{2d})$.\\
For simplicity we assume  here that  $\Theta = \Gamma =i\1$. The general case can be reached 
by the same method or by using the deformation argument
of proof I as we shall see later for more general Hamiltonians.\\
Here the Hamiltonian $\hat H(t)$ is a quadratic form. So using dilations we can assume that $\hbar=1$.  A simple computation 
 left to the  reader, gives the following
\begin{lemma}
\beq
(g^{-1}\hat H(t)g)(x) = Gx\cdot x +i(L+L^\tau)x\cdot x -Kx\cdot x + {\rm Tr}(K-iL)
\eeq
where $g(x) = {\rm e}^{-\frac{\vert x\vert^2}{2}}$.
\end{lemma}
So we get
\beq
(i\partial_t-\hat H(t))\psi(t) = (2\pi\hbar)^{-3d/2}\int\!\!\!\int_{\R^{2d}\times\R^d}{\rm e}^{\frac{i}{\hbar}\Phi^{(\Theta,\Gamma)}(t,z;x,y)}
b(t,x,z)\psi(y)dzdy
\eeq
where 
$$
b(t,z,z) =i\partial_t k(t)  - k(t)\bigl(E(x-q_t)\cdot (x-q_t) + {\rm Tr}(K-iL)\bigr)
$$
As in proof I, we  integrate by parts in the variable $z\in\R^{2d}$, using
$$
(\partial_q -i\partial_p)\Phi = M^\tau(x-q_t)
$$
with $M = C-B -i(A+D)$,  which is invertible (see below Lemma \ref{mip}).  Using the Hamilton  equation of motion we  get
\beq
\dot M = -E(A-iB) -i (K-iL)M.
\eeq
So, we find the following differential equation for $k(t)$,
\beq
\dot{k} = \frac{1}{2}{\rm Tr}(\bigl(M\dot{M}\bigr)k.
\eeq
 Using the Liouville formula,  we get again (\ref{kquad}) for this particular phase. \sq

\section{Proof of Theorem \ref{Thm1} and Theorem \ref{eht}}
As usual for this kind of problems there are  two steps : 1-Determine the amplitudes $a_j$ solving by induction  transport differential equations, 
2-Estimate the error between the approximated propagator  and the exact one.
\subsection{Transport equations}
It is convenient to write 
\beq
{\rm e}^{\frac{i}{\hbar}\Phi} = (\pi\hbar)^{d/2}\varphi_{z_t}(x)\bar\varphi_z(y){\rm e}^{\frac{i}{\hbar}(S(t,z)+(p\cdot q -p_t\cdot q_t)/2)}
\eeq
Then we have to  compute $\hat H^\hbar(t)\varphi_{z_t}$. It is not difficult to add contributions of the lower order terms
of the Hamiltonian, so we shall assume for simplicity that $H^\hbar(t)=H_0(t):=H(t)$.  
\begin{lemma}\label{Tayl}
For every $N\geq 2$ we have
\beq
\hat H(t)\varphi_{z_t}(x) = \sum_{\vert\gamma\vert\leq N}\frac{\hbar^{\vert\gamma\vert/2}}{\gamma!}\partial_X^\gamma H(t,z_t)
\Pi_\gamma\Bigl(\frac{x-q_t}{\sqrt{\hbar}}\Bigr)\varphi_{z_t}(x) + \hbar^{(N+1)/2}T(z_t)\Lambda_\hbar Op^w_1[R_N(t,z_t)]g(x)
\eeq
where
\beq\label{RT}
R_N(t,z_t, X) = \int_0^1\frac{(1-s)^N}{N!}\sum_{\vert\gamma\vert=N+1}\partial_X^\gamma H(t, z_t+s\sqrt\hbar X)X^\gamma ds
\eeq
and $\Pi_\gamma$ is a universal polynomial of degree $\leq\vert\gamma\vert$  which is even or odd according $\vert\gamma\vert$ is even or odd.
\end{lemma}
{\bf Proof.} Let us recall that  $\varphi_z = \hat T(z)\Lambda_\hbar g$. In this proof we put $z_t=z$.
  An easy property of Weyl quantization gives 
 \beq
 \Lambda_\hbar^{-1}\hat T(z)\hat H(t)\hat T(z)\Lambda_\hbar =Op^w_1[H(\sqrt\hbar\bullet +z)]
 \eeq
 So the Lemma follows easily from the Taylor formula with integral remainder.\sq
 
 In this first step we don't take care of remainder estimates, this will be done in the next step.\\
 Let us denote ${\cal I}(a, \Phi)$ the formal operator having the Schwartz kernel
 \beq
 K_a(x,y) = (2\pi\hbar)^{-3d/2}\int_{\R^{2d}}{\rm e}^{\frac{i}{\hbar}\Phi(t,z;x,y)}a(t,z)dz
 \eeq
 From the Lemma \ref{Tayl} we can write
 \bea
 \hat H(t){\cal I}(a,\Phi) \sim {\cal I}(b,\Phi),\;\; {\rm where} \nonumber\\
 b \sim \sum_{\gamma}\frac{\hbar^{\vert\gamma\vert/2}}{\gamma!}\partial_X^\gamma H(t,z_t)
\Pi_\gamma\Bigl(\frac{x-q_t}{\sqrt{\hbar}}\Bigr)a
 \eea
 We have
 \beq
 \Pi_\gamma(x) = \sum_{\beta\leq\gamma}h_{\gamma, \beta}x^\beta
 \eeq
 The quadratic part can be computed as for quadratic Hamiltonians and the linear part disappears with the classical motion. So we have
 \bea\label{trans1}
 b \sim H(t,z_t)a + (\partial_qH(t,z_t) +i\partial_p H(t,z_t))\cdot (x-q_t)a + \nonumber\\
 \hbar\Bigl(E\bigl(\frac{x-q_t}{\sqrt\hbar}\bigr)\cdot\bigl(\frac{x-q_t}{\sqrt\hbar}\bigr) + {\rm Tr}(K-iL)\Bigr)a
 \eea
 where we denote $\partial_{X,X}^2H(t,X)$  the Hessian matrix of $H(t)$.  We have 
 \beq
 \partial_{X,X}^2H(t,z_t) = \begin{pmatrix} G & L\\ L & K\end{pmatrix},\;\; E=G+2iL-K.
 \eeq 
with  $G:=\partial_{q,q}^2H(t,z_t)$, $L:=\partial_{q,p}^2H(t,z_t) $, $K:=\partial_{p,p}^2H(t,z_t)$.\\
Here the stability matrix  $F_t = \begin{pmatrix}A_t & B_t\\C_t&D_t\end{pmatrix}$ satisfies $\dot F_t = J\partial_{X,X}^2H(t,z_t)F_t$, $F_{t=0}=\1$.\\
As in the quadratic case we want to transform the power of $(x-q_t)$ into power of $\hbar$.
\begin{lemma}\label{mip}
Let us denote $M_t =(C_t-B_t)-i(A_t+D_t)$. We have
\beq\label{min}
\bigl\vert\det M_t\bigr\vert \geq 2^{-d},\;\; {\rm and}
\eeq
\beq\label{ipp}
\hbar(\partial_q-i\partial_p){\rm e}^{\frac{i}{\hbar}\Phi} = iM_t^\tau(x-q_t){\rm e}^{\frac{i}{\hbar}\Phi}
\eeq
\end{lemma}
{\bf Proof.}  For simplicity, let us forget the lower index $t$. \\
Let us consider the $2d\times 2d$ matrix
\beq
\1+F+iJ(\1-F) = \begin{pmatrix}\1+A-iC & B+i(\1-D)\\ C-i(\1-A) & \1+D+iB\end{pmatrix} =  
\begin{pmatrix}\1+A-iC & -i(D+iB) +i\\ i(A-iC) & \1+D+iB\end{pmatrix}
\eeq
Using the Lemma 4 in \cite{fo}, Appendix A, we get
\beq
\det(\1+F+iJ(\1-F)) = \det\bigl((\1+A-iC)( \1+D+iB)-(A-iC-\1)(D+iB-\1)\bigr)= 2^d\det\big(A+D+i(B-C)\bigr)
\eeq
Using that $F$ is symplectic, we get
\beq
(\1+F+iJ(\1-F))^*(\1+F+iJ(\1-F)) = (\1+F^\tau)(\1+F) + (1-F^\tau)(\1-F) \geq \1_{2d}
\eeq
hence (\ref{min}) follows.\\
Let us recall classical computations for the derivatives of the action
\bea\label{daction}
\partial_q S &=& (\partial_qq_t)^\tau p_t -p \\
\partial_p S &= & (\partial_pq_t)^\tau p_t
\eea
Then we can compute $\partial_q\Phi$, $\partial_p\Phi$ and we get (\ref{ipp}). \sq

Integrate by parts like in the quadratic case, we get
\beq
(i\hbar\partial_t-\hat H(t)){\cal I}(a, \Phi) \asymp {\cal I}(f, \Phi)
\eeq
where
\bea
 f \sim i\hbar\bigl(\partial_t a - \frac{1}{2}{\rm Tr}\bigl(\dot M M^{-1}\bigr)a \nonumber \\
+  \sum_{\vert\gamma\vert\geq 3}\frac{\hbar^{\vert\gamma\vert/2}}{\gamma!}\partial_X^\gamma H(t,z_t)
\Pi_\gamma\Bigl(\frac{x-q_t}{\sqrt{\hbar}}\Bigr)a
\eea
Hence using the Liouville formula, we get the first term
\beq
a_0(t, z) = 2^{d/2}{\rm det}^{1/2}\bigl(iM\bigr)
\eeq
We shall obtain the next terms $a_j$  by  successive integrations by parts.  This is solved  more explicitly with the following Lemma.
\begin{lemma}
For any  symbol $b\in{\cal O}_0(2d)$, and every multiindex $\alpha\in\N^{2d}$ we have
\beq\label{conv}
\int_{\R^{2d}}(x-q_t)^\alpha{\rm e}^{\frac{i}{\hbar}\Phi} b(z)dz =
\sum_{\frac{\vert\alpha\vert}{2}\leq\vert\beta\vert\leq\vert\alpha\vert} \hbar^{\vert\beta\vert}
\int_{\R^{2d}}f_{\alpha,\beta}(t,z){\rm e}^{\frac{i}{\hbar}\Phi}\partial_z^\beta b(z)dz
\eeq
where $f_{\alpha,\beta}(t,z)$ are symbols of order 0, uniformly bounded in ${\cal O}_0(2d)$ on bounded time intervals.
They only depend on the classical flow $\phi^t(z)$ and its derivatives.\\
More precisely, let us assume that there exists a non positive function $\mu(T)$ such that for every
$\gamma\in \N^{2d}$ we have
\beq
\sup_{\vert a \vert\leq T}\vert\partial_z^\gamma\phi^t(z)\vert \leq C_\gamma\mu(T)^{\vert\gamma\vert}
\eeq
Then we have
\beq
\vert\partial_z^\epsilon f_{\alpha,\beta}(z)\vert \leq C_{\alpha,\beta;\epsilon}\mu(T)^{\vert\alpha\vert-\vert\beta\vert+\vert\epsilon\vert}
\eeq
\end{lemma} 
{\bf Proof.}  The Lemma is easily obtained by induction on $\vert\alpha\vert$ using Lemma \ref{mip} \sq

Now, to determine the transport equation,  we solve inductively on $j\geq 0$, the equation
\beq\label{dtr}
(i\hbar\partial_t-\hat H(t)){\cal I}\Bigl(\sum_{0\leq k\leq j+1}\hbar^ka_k(t), \Phi\Bigr) = O(\hbar^{j+2})
\eeq
Reasoning by induction on $j\geq 0$,  we get the transport equation   for $a_{j+1}(t)$ by cancellation of the coefficient of
$\hbar^{j+1}$  in (\ref{dtr}). 
\beq\label{trj}
\partial_t a_{j+1}(t,z) = \frac{1}{2}{\rm Tr}\bigl(\dot M M^{-1}\bigr)a_{j+1}(t,z) + b_j(t,z),\;\;\; a_{j+1}(0,z)= 0,
\eeq
where
\beq\label{rtrj1}
b_j(t,z) = \sum_{\vert\alpha\vert +2k \leq 2(j+2)}F_{j,k,\alpha}(t,z)\partial_z^\alpha a_k(t,z).
\eeq
Moreover, $F_{j,k,\alpha}(t,z)$ depends only on the classical flow $\phi^t(z)$ and its derivatives and satisfies
\beq\label{rtrj2}
\vert\partial_z^\gamma F_{j,k,\alpha}(t,z)\vert \leq C_{j,k,\alpha,\gamma}\mu(T)^{2(j-k+2)+\vert\gamma\vert -\vert\alpha\vert}
\eeq
where $ C_{j,k,\alpha,\gamma}$ only depends on 
$\di{\sup_{\vert t\vert\leq T}\vert H(t)\vert_{\infty,\gamma}}$,  $2\leq\vert\gamma\vert\leq j+2$.\\
So we get, for every $j\geq 0$,
\beq\label{ftrj}
a_{j+1}(t,z) = \int_0^t{\rm det}^{1/2}\bigl(M(t,z)M(s,z)^{-1}\bigr)b_j(s,z)ds
\eeq
Moreover, from (\ref{rtrj1}) and (\ref{rtrj2}), we get the following estimate, for every $j\geq 0$, $\vert t\vert \leq T$, $z\in\R^{2d}$,
\beq\label{estaj}
\vert\partial_z^\gamma a_j(t,z)\vert \leq C_{j,\gamma}\vert{\rm det}^{1/2}M(t,z)\vert\mu(T)^{4j +\vert\gamma\vert}
\eeq
with the same remark as in (\ref{rtrj2}) for the constant $C_{j,\gamma}$.
\subsection{Error estimates}
Let us denote 
\beq
{\cal R}_N(t) = (i\hbar\partial_t-\hat H(t)){\cal I}\Bigl(a^{(N)}(t), \Phi\Bigr)
\eeq
where $\di{a^{(N)}(t) = \sum_{0\leq k\leq N}\hbar^ka_k}$.  Using Duhamel formula  we have
\beq
\Vert U^{\hbar}(t) - U^{N,\hbar}(t)\Vert \leq \hbar^{-1}\int_0^t\Vert{\cal R}(s)\Vert ds
\eeq
where $t_0=0$,  $U^{\hbar}(t) = U^{\hbar}(t,0)$, $U^{N,\hbar}(t) ={\cal I}\Bigl(a^{(N)}(t), \Phi\Bigr)$.\\
So we have to estimate $\Vert{\cal R}_N(t)\Vert$.  Let us denote $ K^{(N)}(x,y)$ the Schwartz kernel of ${\cal R}_N(t)$
and $\tilde K^{(N)}(X,Y)$ the Schwartz kernel of ${\cal R}_N(t)$ in the Fourier-Barmann representation :
\beq
\tilde K^{(N)}(X,Y) = \int\!\!\!\int_{\R^d\times\R^d} K^{(N)}(x,y)\varphi_X(y)\overline{\varphi_Y(x)}dxdy.
\eeq
Let be $\tilde{\cal R}_N(t)$ the operator with Schwartz kernel $\tilde K^{(N)}(X,Y)$. The following Lemma is well known.
Here we forget $N$ and $t$ for simplicity. 
\begin{lemma}\label{Rest}
We have the $L^2$ norm estimate
\beq\label{barest}
\Vert{\cal R}\Vert_{L^2(\R^d)} \leq (2\pi\hbar)^{-d}\Vert\tilde{\cal R}\Vert_{L^2(\R^d)}
\eeq
In particular we have 
\bea\label{Carl}
\Vert{\cal R}\Vert_{L^2(\R^d)} \leq (2\pi\hbar)^{-d}
\max\Big\lbrace\sup_Y\int\vert\tilde K(X,Y)\vert dX, \sup_X\int\vert\tilde K(X,Y)\vert dY\Big\rbrace
\eea
\end{lemma}
{\bf Proof } For inequality (\ref{barest})  we use that  the Fourier-Bargmann transform is an isometry.\\
Inequality (\ref{Carl}) is known as Carleman (or Schur) $L^2$ estimate.\sq \\
Using Lemma \ref{Tayl} we get 
\bea
\tilde K^{(N)}(X,Y) = 2^{-3d/2}(\pi\hbar)^{-d}\int_{\R^{2d}}\bigl\langle\hat T (z_t)\Lambda_\hbar Op^w_1[R_N(t)]g,\varphi_Y\bigr\rangle
\langle\varphi_X,\varphi_z\rangle a^{(N)}(t,z){\rm e}^{\frac{i}{\hbar}\delta(t,z}dz
\eea
where $\delta(t,z) = S(t,z) +\frac{p\cdot q - p_t\cdot q_t}{2}$.\\
Using Weyl commutation formula  we have
\bea
\langle\varphi_X,\varphi_z\rangle &=& \exp\Bigl(-\frac{\vert X-z\vert^2}{4\hbar} +\frac{i}{2\hbar}\sigma(X,z)\Bigr)\\
\bigl\langle\hat T (z_t)\Lambda_\hbar Op^w_1[R_N(t)]g,\varphi_Y\bigr\rangle &=& 
\bigl\langle Op^w_1[R_N(t)]g,g_{\frac{Y-z_t}{\sqrt\hbar}}\bigr\rangle.
\eea
We know   the  Wigner function $W_{0,Z}$ of the pair $(g,g_Z)$, $Z\in\R^{2d}$ \cite{ro2}
\beq
W_{0,Z}(X) = 2^{2d}\exp\Bigl(-\vert X-\frac{Z}{2}\vert^2 - i\sigma(X,Z)\Bigr)
\eeq
By a well known  property of Weyl quantization \cite{fo}, for any symbol $s$, we have
\beq
\langle Op^w_1[s]g,g_Z\rangle = (2\pi)^{-d}\int_{\R^{2d}}s(X)W_{0,Z}(X)dX
\eeq
We shall use the following Lemma
\begin{lemma}\label{estst}
Let be $f\in{\cal O}_0(2d)$.  For every $\gamma\in\N^{2d}$  and $m>0$ there exists $C_{\gamma, m}$ such that
\beq
\Bigl\vert\int_{\R^{2d}} X^\gamma f(X){\rm e}^{-\vert X-Z\vert^2 -iJZ\cdot X}dX\Bigr\vert
\leq C_{\gamma, m}(1+\vert Z\vert)^{-m} \sup_{\vert\alpha\vert\leq m+\vert\gamma\vert;\; Y\in\R^{2d}}\vert\partial_Y^\alpha f(Y)\vert
\eeq
\end{lemma}
{\bf Proof} It is enough to assume $\vert Z\vert \geq 1$. We integrate $m$ times by parts with the differential operator
\beq
{\cal L} = \frac{2(X-Z) -iJZ\cdot X}{4\vert X-z\vert^2 + \vert JZ\vert^2}\partial_X
\eeq
using that $\di{({\cal L}^\tau)^m = \sum_{\vert\alpha\vert\leq m}l_{m,\alpha}\partial_X^\alpha}$, with
$\vert l_{m,\alpha}\vert \leq C_{m,\alpha}(\vert Z\vert + \vert X-Z\vert)^{-m}$, 
where  \\ $\theta(X) = -\vert X-Z\vert^2 -iJZ\cdot X$. \sq\\
So using Lemma \ref{estst} we get the following estimate~: for every $N;N^\prime$ there exists $C_{N,N^\prime}$  (depending only on 
semi-norms  $\vert H(t)\vert_{\infty,\gamma}$, $2\leq\vert\gamma\vert \leq N+N^\prime$, 
 such  that for $X,Y\in\R^{2d}$ and $\vert t\vert \leq T$ we have 
\beq\label{estbarg}
\vert\tilde K^{(N)}(X,Y)\vert \leq C_{N,N^\prime}(\mu(T))^{N+N^\prime}\hbar^{\frac{N+1}{2}-d}\int_{\R^{2d}}{\rm e}^{-\frac{\vert X-z\vert^2}{4\hbar}}
\Bigl(1+\frac{\vert Y-z_t\vert}{\sqrt\hbar}\Bigr)^{-N^\prime}\vert a^{(N)}(t,z)\vert dz.
\eeq
Let us denote $ \phi^{*t}=\phi^{0,t}=(\phi^t)^{-1}$. We have the Lipchitz estimate, for $\vert t\vert \leq T$, 
\beq
\vert\phi^{*,t}Y-z\vert \leq \mu(T)\vert Y-z_t\vert 
\eeq
So we get 
\beq
\Bigl\vert\int_{\R^{2d}}{\rm e}^{-\frac{\vert X-z\vert^2}{4\hbar}}
\Bigl(1+\frac{\vert Y-z_t\vert}{\sqrt\hbar}\Bigr)^{-N^\prime}dz\Bigr\vert 
\leq C_{N^\prime} \Bigl(1+ \frac{\vert{\phi^t}^*Y-X\vert}{\mu(T)\sqrt\hbar}\Bigr)^{-N^\prime}
\eeq
and 
\beq\label{FIO}
\vert\tilde K^{(N)}(X,Y)\vert \leq C_{N,N^\prime}(\mu(T))^{N+N^\prime}\hbar^{\frac{N+1}{2}}\Bigl(1+ \frac{\vert{\phi^t}^*Y-X\vert}{\mu(T)\sqrt\hbar}\Bigr)^{-N^\prime}\sup_{z\in\R^{2d}, \vert t \vert\leq T}\vert a^{(N)}(t,z)\vert
\eeq
Then using Lemma \ref{Rest} and choosing $N^\prime >2d$, we get the following uniform $L^2$ estimate for the remainder term, 
for $\vert t\vert \leq T$, 
\beq
\Vert{\cal R}_N(t)\Vert \leq C_{N}(\mu(T))^{N+1}\hbar^{(N+1)/2}\sup_{z\in\R^{2d}, \vert t \vert\leq T}\vert a^{(N)}(t,z)\vert
\eeq
If $T$ is fixed, pushing the expansion up to $2N$ instead of $N$ we get easily Theorem \ref{Thm1} using Duhamel formula.\\
Using  global estimates on $a_j(t,z)$ obtained  from the transport equation (\ref{estaj})  and pushing the asymptotic expansion up to $2N$,
we get  the proof of Theorem \ref{eht} using again  Duhamel formula.

\section{Varying  phase. Proof of Theorem \ref{PVVK}}
To avoid  technicalities we fix the time $t$. It would be not difficult to follow  a time parameter $t$ if necessary for application. 
So in this section $\phi$ is a symplectic diffeomorphism
in $\R^{2d}$, such that $\phi$, $\phi^{-1}$ are Lipchitz continuous and $\phi\in{\cal O}_1(2d)$.\\
We denote  $z=(q,p)\in\R^{2d}$, $\phi(z) = (Q(z), P(z))\in\R^d\times\R^d$ and $S$ an action for $\phi$, i.e a primitive on $\R^{2d}$  of the closed 1-form
$PdQ-pdq$. We consider the following  phases
\beq
 \Phi^{(\phi,\Theta,\Gamma)}(z;x,y) = S(z) + P\cdot(x-Q) -p\cdot(y-q) +\frac{1}{2}
\bigl(\Theta(x-Q)\cdot(x-Q) - \overline{\Gamma}(y -q).(y-q)\bigr)
\eeq
This class of Fourier-Integral  operators with complex quadratic  phase was   ready analyzed in \cite{rosw}. We want here to show how to vary
 the choice of  the matrices $\Theta, \Gamma$
for a given canonical transformation $\phi$ of $\R^{2d}$.  As in section 3, let us denote ${\cal I}(a,\Phi)$ the operator with the Schwartz kernel
\beq
 K_a(x,y) = (2\pi\hbar)^{-3d/2}\int_{\R^{2d}}{\rm e}^{\frac{i}{\hbar} \Phi^{(\phi,\Theta,\Gamma)}(z;x,y)}a(z)dz
 \eeq
 where $a\in{\cal O}_0(2d)$, $\Phi = \Phi^{(\phi,\Theta, \Gamma)}.$ \\
 Using a Fourier-Bargmann transform  and the following estimate~: there exist $C>0$, $c>0$ such that for all $X\in\R^{2d}$ we have
 \beq\label{scalg}
 \vert\langle\varphi^\Gamma, \varphi_X\rangle\vert \leq C\exp\Bigl(-\frac{c\vert X\vert^2}{\hbar}\Bigr),
 \eeq
 we can estimate the Fourier-Bargmann transform $\tilde K_a(X,Y)$ of $K_a$  and prove that ${\cal I}(a,\Phi)$ is bounded in $L^2(\R^d)$
 (see   section 3, Lemma \ref{estst} and the  section 5 below). \\
 Our  goal in this section is to prove the following result which  gives Theorem \ref{PVVK} as a particular case. 
 \begin{proposition}\label{vary}
 Let be  4 matrices in $\Sigma_+(d)$, $\Theta, \Theta^\prime,\Gamma, \Gamma^\prime$ and $a\in{\cal O}_0(2d)$.
 $\Theta, \Theta^\prime$ may be $z$ dependent such that
 \beq\label{estmat3}
\exists c>0,\; \Im\Theta^{(\prime)} v.v \geq c\vert v\vert^2,\;\;  \forall z\in\R^{2d}
\eeq
\beq\label{estmat4}
\forall\gamma, \vert\gamma\vert \geq 1, \exists C_{\gamma},\; \Vert\partial_z^\gamma\Theta^{(\prime)}\Vert \leq C_{\gamma},\; \forall z\in\R^{2d}. 
\eeq
 Then there exists a semi-classical symbol $a^\prime\sim \sum_j \hbar^j a^\prime_j$ of order 0 such that
 we have for  the $L^2$ operator norm, 
\beq
 {\cal I}(a,  \Phi^{(\phi,\Theta,\Gamma)}) =  {\cal I}(a^\prime,  \Phi^{(\phi,\Theta^\prime,\Gamma^\prime)}) + O(\hbar^\infty)
 \eeq
 Moreover we have for the principal symbol $a^\prime_0$ the formula
 \beq
 a^\prime_0(z) = a_0(z)\frac{{\rm det}^{1/2}(M(1))}{{\rm det}^{1/2}(M(0))}
 \eeq
 where $M(s) :=  C+D\bar\Gamma - \Bigl((1-s)\Theta+s\Theta^\prime\Bigr)(A+B\bar\Gamma)$
 \end{proposition}
 {\bf Proof.} 
 The method is rather simple and is  an extension of what we have already done for quadratic Hamiltonians (Proof I)
 except  that here we have to solve transport equations in the deformation parameter $s$  to get the lower order  correction terms.\\
Let us remark that this class of Fourier-integral operators is closed under adjointness~:
\beq
{\cal I}(a,  \Phi^{(\Theta,\Gamma)})^* = {\cal I}(a^*,  \Phi^*),
\eeq
where $ a^*(Z)=\bar a(\phi^{-1}Z)$,  $ Z=(Q, P)$, $Z=\phi(z)$  and 
\bea
\Phi^*(Z;x,y) =  -S(\phi^{-1}Z) + p\cdot(x-q) -P\cdot(y-Q) + \nonumber\\
\frac{1}{2}
\bigl(\Gamma(x-q)\cdot(x-q) - \overline{\Theta}(y -Q).(y-Q)\bigr)
\eea
So by transitivity we can assume
that $\Gamma = \Gamma^\prime$.  As in the quadratic Hamiltonian case let us introduce 
$\Theta_s = (1-s)\Theta + s\Theta^\prime$, $\Phi^{(s)} = \Phi^{(\Theta_s,\Gamma)}$, $0\leq s\leq 1$   and  look for a semiclassical
symbol $a^{(s)} = \sum_j \hbar^j a^{(s)}_j$ such that 
\beq\label{varasymp}
\frac{\partial}{\partial s}\int_{\R^{2d}}{\rm e}^{\frac{i}{\hbar}\Phi^{(s)}(z;x,y)}a^{(s)}(z)dz = O(\hbar^\infty) ,\;\;\forall s\in[0, 1]
\eeq
But we have
\beq
\frac{\partial}{\partial s}\Phi^{(s)}(z;x,y) = \frac{i}{\hbar}(\Theta^\prime -\Theta)(x-Q)\cdot(x-Q)
\eeq
and we have to find a $C^1$ family symbol $a^{(s)}$,  $0\leq s\leq 1$ such that
\beq
{\cal I}\Bigl(\partial_s a^{(s)} + \frac{i}{\hbar}(\Theta^\prime-\Theta)(x-Q)\cdot(x-Q)a^{(s)}, \Phi\Bigr) = O(\hbar^\infty)
\eeq
The principal  term $a^\prime_0 = a^{(1)}$ is computed  as in the quadratic case.\\ 
Let us suppose for a moment that $\Theta, \Theta^\prime$ are constant. Then 
as in the quadratic case we have
\beq
(\partial_q +\bar\Gamma\partial_p)\Phi^{(s)} = 
\bigl(C^\tau +\bar\Gamma D^\tau - (A^\tau +\bar\Gamma B^\tau)\Theta_s\bigr)(x-Q) 
\eeq
where  $A=\partial_q Q$, $B=\partial_p Q$, $C=\partial_q P$, $D=\partial_p P$
 and $F=\begin{pmatrix} A & B\\ C & D \end{pmatrix}$ is a symplectic matrix.\\
We know that 
$  M(s) :=  C+D\bar\Gamma - \Theta_s(A+B\bar\Gamma)$ is invertible so we can integrate by parts as in section 3.
and as above we can achieve the proof of Proposition \ref{vary}.\\
When $\Theta, \Theta^\prime$ are $z$ dependent, the integrations by part are more tricky. We have to use
\beq\label{ippit1}
(\partial_q +\bar\Gamma\partial_p)\Phi^{(s)} = 
M\tau(s,z)(x-Q)  + N(s,z)(x-Q, x-Q)
\eeq
where $N(s,z)(x,y) $ is a bilinear  application in  $(x,y)\in\R^d\times\R^d$ into $d\times d$ matrices, with coefficients in ${\cal O}_0$
in $z$, $C^1$ in $s$.\\
    Hence we have
    \beq\label{ippt2}
(x-Q){\rm e}^{\frac{i}{\hbar}\Phi^{(\Theta,\Gamma)}} = 
\frac{\hbar}{i}(M^\tau)^{-1}(s,z)\bigl(\partial_q +\bar\Gamma\partial_p\bigr){\rm e}^{\frac{\hbar}{i}\Phi^{(\Theta,\Gamma)}}
 - M^\tau)^{-}(s,z)N(s,z)(x-Q,x-Q){\rm e}^{\frac{\hbar}{i}\Phi^{(\Theta,\Gamma)}}
\eeq
So we apply  (\ref{ippt2}) and the following lemmas to proceed like in  section 3.
\begin{lemma}\label{trest}
For any  symbol $b\in{\cal O}_0(2d)$, for every multiindex $\alpha\in\N^{2d}$ and every $N\geq\vert\alpha\vert/2$  we have
\bea\label{conv2}
\int_{\R^{2d}}(x-Q)^\alpha{\rm e}^{\frac{i}{\hbar}\Phi^{(s)}} b(z)dz =
\sum_{\frac{\vert\alpha\vert}{2}\leq\vert\beta\vert\leq N } \hbar^{\vert\beta\vert}
\int_{\R^{2d}}f_{\alpha,\beta}(s,z){\rm e}^{\frac{i}{\hbar}\Phi^{(s)}}\partial^\beta b(z)dz + \nonumber\\
\sum_{\vert\beta\vert +\vert\gamma\vert = N+1, \vert\beta\vert\geq 1}\hbar^{\vert\gamma\vert}
\int_{\R^{2d}}g_{\alpha,\beta}(s,z)(x-Q)^\beta{\rm e}^{\frac{i}{\hbar}\Phi^{(s)}}g_{\beta,\gamma}\partial^\gamma b(z)dz
\eea
where $f_{\alpha,\beta}(s,z)$, $g_{\alpha,\beta}(s,z)$  are symbols of order 0, uniformly bounded in ${\cal O}_0(2d)$ for $s\in[0,1]$. 
\end{lemma}
\begin{lemma}\label{restbd}
For every $b\in{\cal O}_0(2d)$ and $\beta\in\N^d$ we have the crude $L^2$ estimate, uniform in $s\in[0,1]$, 
\beq
\Vert{\cal I}((x-Q)^\beta b, \Phi^{(s)}\Vert = O\bigl(\hbar^{\vert\beta\vert/2}\bigr)
\eeq
\end{lemma}
Using these two lemmas we get  the full semiclassical symbol $a^\prime\sim \sum_j \hbar^j a^\prime_j$, where 
\beq
a^\prime_0(z) = a_0\frac{{\rm det}^{1/2}(M(s))}{{\rm det}^{1/2}(M(0))}
\eeq
and for $j\geq 1$,  $a^\prime_j$ is computed by induction as solution for $s=1$ of the differential equation
\beq
\partial_s a_j(s) = {\rm Tr}\bigl(\dot M(s)M^{-1}(s)\bigr)a_j(s) + b_j(s),\;\; a_j(0)=a_j.
\eeq
where $b_j(s) $ depends on the $a_k(s)$, $k\leq j-1$ \sq
\begin{remark}
Considering the adjoint operator, it is possible to exchange the role of the matrices $\Theta$ and $\Gamma$.\\
If the symbol $a$ depends smoothly on some parameter $\lambda$, it is not difficult to show that $a^\prime$
also depends smoothly in $\lambda$.
\end{remark}
{\bf Proof of Lemma \ref{trest}.} This is  done by an induction on $N$ such that $\alpha\leq N$.\sq\\
{\bf Proof of Lemma \ref{restbd}.}   Let us begin by giving a simple proof of (\ref{scalg}) when $\Theta$
is $z$ dependent satisfying the assumptions (\ref{vary}). We shall prove the more general estimate,
for  every $\beta\in \N^d$ there exist $C>0$, $c>0$ such that
\beq\label{polyscal}
\vert\langle x^\alpha g^\Theta, g_Y\rangle\vert  \leq C{\rm e}^{-c\vert Y\vert^2},\;\; \forall Y\in\R^{2d}
\eeq
Let us denote $Y=(y,\eta)\in\R^d\times\R^d$. 
By a direct estimate we get easily,
\beq
\vert\langle x^\alpha g^\Theta, g_Y\rangle\vert \leq C{\rm e}^{-2c\vert y\vert^2},\;\; \forall (y,\eta)\in\R^{2d}.
\eeq
Using  Fourier transform  and Plancherel formula, we exchange $y$ and $\eta$ and we get (\ref{polyscal}).\\
Now we can follow the method of section 3 to estimate $L^2$ norm of operators using a  Fourier-Bargmann transformation.\\
Let be $\tilde K(X,Y)$ the Fourier-Bargmann kernel of ${\cal I}((x-Q)^\beta b, \Phi^{(s)})$. We have
\bea
\tilde K(X,Y) = 2^{-3d/2}(\pi\hbar)^{-d}\hbar^{\vert\alpha\vert/2}\int_{\R^{2d}}\bigl\langle\hat T(Z)\Lambda_\hbar(x^\beta g^\Theta),\varphi_Y\bigr\rangle
\langle\varphi_X,\varphi_z\rangle b(z){\rm e}^{\frac{i}{\hbar}\delta(t,z)}dz
\eea
 where $Z=(Q,P)=\phi(z)$ and
\beq
\vert\bigl\langle\hat T(Z)\Lambda_\hbar(x^\beta g^\Theta),\varphi_Y\bigr\rangle\vert = 
\vert\langle x^\beta g^\Theta, g_{\frac{Y-Z}{\sqrt\hbar}}\rangle\vert.
\eeq
So we get
\beq
\vert\tilde K(X,Y)\vert \leq C\hbar^{\vert\alpha\vert/2}\int_{\R^{2d}}\exp\Bigl(-\frac{c}{\hbar}(\vert Y-\phi(z)\vert^2 +\vert X-z\vert^2\Bigr)dz
\eeq
Using that $\phi$ is  a Lipchitz canonical transformation, we have, for $C_0$ large enough and $c_0>0$ small enough,
\beq
\vert\tilde K(X,Y)\vert \leq C_0\hbar^{\vert\alpha\vert/2}\exp\Bigl(-\frac{c_0}{\hbar}(\vert Y-\phi(X)\vert^2\Bigr)
\eeq
Hence we get the proof of Lemma \ref{restbd} using Lemma \ref{Rest}.\sq \\
 we have proved Proposition \ref{vary} and Theorem  \ref{PVVK}. 

\section{Semiclassical Fourier Integral Operators}
In \cite{lasi},  \cite{bu} and in  the recent preprint \cite{ro},  the authors have considered Fourier-integral operators defined by the following 
simpler phase
\beq\label{lasiph}
\Psi^{(\phi,\Theta)}(p;x,y) = S(y,p) + P\cdot(x-Q)  +\frac{1}{2}\Theta(x-Q)\cdot(x-Q) 
\eeq
where $(Q,P) =\phi(y,p)$, $\phi$ is a  bilipchitz canonical transformation like above, $\Theta\in\Sigma_+(d)$.\\
In \cite{lasi} and  \cite{bu}  the authors  have proved semiclassical expansions  for the propagator of Schr\"odinger equation for initial data with a compact support.
This result is extended  in \cite{ro}  for the Schr\"odinger Hamiltonian $-\hbar^2\triangle +V$, to general data in $L^2$   with uniform 
 norm estimates. We shall  give here some extensions  of results of  \cite{ro}  using the same techniques as in section 3 and 4, so we shall not
  repeat the details.

Let us denote ${\cal J}(a,\Psi^{\phi,\Theta})$  the operator whose  Schwartz kernel is 
\beq
K(x,y)= (2\pi\hbar)^{-d}\int_{\R^d}{\rm e}^{\frac{i}{\hbar}\Psi^{(\phi,\Theta)}(p;x,y)}a(y,p)dp
\eeq
A natural question discussed in this section is to compare the Fourier-Integral operators ${\cal I}(a, \Phi^{(\phi,\Theta, \Gamma)})$ 
defined with $2d$ ``frequency variables" and ${\cal J}(a,\Psi^{(\phi,\Theta)})$ defined with $d$ ``frequency variables".\\
A Fourier integral operator  in $L^2(\R^d)$ is always  a quantization  of a canonical transformation  $\phi$ in the cotangent
space $T^*(\R^d)$.  A nice way to make clear this relationship is to use a Fourier-Bargmann transform (see \cite{bo, ta}). 
This can be  easily done   in the same way for Semiclassical-Fourier-Integral operators as we shall see now.
\begin{definition}\label{defio}
A family of operators, depending on a small parameter $\hbar\in]0, 1]$,  $U^\hbar: {\cal S}(\R^d) \rightarrow {\cal S}^\prime(\R^d)$ 
is a Semiclassical-Fourier-Integral operator of order $m\in\R$
associated to the canonical bilipchitz transformation $\phi: T^*(\R^d)\rightarrow T^*(\R^d)$, if for every $N^\prime$ 
we have $U^\hbar = U_{N^\prime}^\hbar + R^\hbar_{N^\prime}$ where 
$U_{N^\prime}^\hbar : {\cal S}(\R^d) \rightarrow {\cal S}^\prime(\R^d)$    and $\Vert R^\hbar_{N^\prime}\Vert  = O(\hbar^{N^\prime})$
and for every $N\geq 0$ there exists $C_N$ such that 
\beq
\vert\tilde K^\hbar(X,Y)\vert \leq  C_N\hbar^{m-3d/2}\Bigl(1+\frac{\vert Y-\phi(X)\vert}{\sqrt\hbar}\Bigr)^{-N},\;\; \forall X, Y\in \R^{2d},\;\; \hbar\in]0, 1],
\eeq
where $\tilde K^\hbar(X,Y)$ is the Schwartz kernel of ${\cal F}_{\cal B}U_{N^\prime}^\hbar{\cal F}^*_{\cal B}$.
\end{definition}
\begin{remark}
\begin{enumerate}
\item In this definition,  which co\"{\i}ncides with a definition given in \cite{ta} for $\hbar=1$,  a semiclassical-Fourier-Integral operator  has,  up to a negligible  operator in  $\hbar$,  a kernel living in a neighborhood
of the graph of a canonical transformation $\phi$. But this definition says nothing concerning asymptotic expansion of $\tilde K^\hbar(X,Y)$
in a neighborhood of the graph of $\phi$ when $\hbar$ is small. So this definition is certainly too permissive. But for $\hbar$ fixed  it is 
suitable as proven in \cite{ta}.
\item
Using  Carleman-Schur estimate, a semiclassical F.I.O of order  0 is uniformly bounded in $L^2(\R^d)$.
This is a straightforward  consequence of the definition. This class of  semiclassical F.I.O of order  0 is clearly
closed by composition.
\item
In  {\rm Definition} \ref{defio}  it is equivalent  to use any Fourier-Bargmann transformation 
${\cal F}^{(\Gamma)}_{\cal B}$, $\Gamma\in\Sigma_+(d)$.
\item
There are other definitions of semiclassical F.I.O using Lagrangian analysis and real phase functions. For this point of view see for example \cite{al}.
\item
Fourier-Integral operators with complex phase  were  used to study propagation of singularities of P.D.E. 
Many  papers and books  have been published on this subject, among them let us point out \cite{babu, ra, sj}.
\end{enumerate}
\end{remark}
Now we  shall see that the operators  already considered in this paper are  semiclassical-Fourier-Integral operators.
\begin{proposition}
Let be amplitudes $a=a(x,z)$, $a\in {\cal O}_0(3d)$   and $u = u(x,y, p)$, $u\in {\cal O}_0(3d)$
and $\Theta, \Gamma \in \Sigma_+(d)$, $\Theta$ may depend in $z$ or $(y,p)$, such that  (\ref{estmat1}), (\ref{estmat2})
are satisfied. Then ${\cal I}(a,  \Phi^{(\phi,\Theta,\Gamma)})$ and ${\cal J}(u,\Psi^{\phi,\Theta})$
are semiclassical-Fourier-Integral operators of order 0.
\end{proposition}
{\bf Proof.} 
Concerning  ${\cal I}(a,  \Phi^{(\phi,\Theta,\Gamma)})$, we get the result following  subsection 3.2, estimate (\ref{FIO}).\\
The proof for ${\cal J}(u,\Psi^{\phi,\Theta})$ is almost the same. For simplicity we assume $\Theta$ constant.  For $\Theta$ depending
in $(y,p)$ we could proceed as in section 4.\\
Let us denote $X=(\tilde x,\tilde\xi)$, $Y=(\tilde y, \tilde\eta)$. We want to estimate 
\beq
\tilde K(X,Y)=(2\pi\hbar)^{-d}\int\!\!\int\!\!\int_{\R^{3d}}{\rm e}^{\frac{i}{\hbar}\tilde\Phi}u(x,y,p)dpdxdy
\eeq
where 
\bea
\tilde\Phi = S(y,p) + P\cdot(x-Q) +\frac{\Theta}{2}(x-Q)\cdot(x-Q) + \nonumber\\
 \frac{i}{2}(\tilde x-y)\cdot(\tilde x-y) +\tilde\xi\cdot(\tilde x-y) 
+ \frac{i}{2}(\tilde y-x)\cdot(\tilde y-x) +\tilde\eta\cdot(\tilde y-x)
 \eea
 Let us remark that we have~:
$F^{-1} = \begin{pmatrix} D^\tau & -B^\tau\\ -C^\tau & A^\tau \end{pmatrix}$
if $ F= \begin{pmatrix} A &  B\\ C & D\end{pmatrix}$. So, because $F^{-1}$ is symplectic,  we know that $D^\tau-B^\tau\Theta$ is invertible.
Hence we  have
\bea
\partial_y\tilde\Phi &=& (C^\tau -A^\tau\Theta)(x-Q) + (\tilde\xi-p) +i(\tilde x-y)\\
\partial_p\tilde\Phi &=& (D^\tau -B^\tau\Theta)(x-Q)\\
\partial_x\tilde\Phi &=& \Theta(x-Q) +(P-\tilde\eta) + i(\tilde y-x).
\eea
So we get the necessary estimates on $\tilde K$ by  integrations by parts using
\bea
\partial_y\tilde\Phi - (-A^\tau\Theta +C^\tau)(D^\tau -B^\tau\Theta )^{-1}\partial_p\Psi &=  (\tilde\xi-p) +i(\tilde x-y)\\
\partial_x\tilde\Phi -\Theta(D^\tau -B^\tau\Theta )^{-1}\partial_p\Psi &= (P-\tilde\eta) + i(\tilde y-x)
\eea
\hfill \sq

The following result is a slight generalization of \cite{lasi, bu, ro}.
\begin{theorem}\label{lasibu}
Under the assumptions of Theorem \ref{Thm1} and  (\ref{estmat1}), (\ref{estmat2}), we have 
\beq\label{VVK2}
K(t;x,y) \asymp (2\pi\hbar)^{-d}\int_{\R^{d}}{\rm e}^{\frac{i}{\hbar}\psi^{(\phi^t,\Theta_t)}(t,y,p,x)}u(\hbar;t,y,p)dp
\eeq
where  $\di{u(\hbar;t,y,p) =\sum_{0\leq j<+\infty}u_j(t;y,p)\hbar^j}$ has the same meaning as in Theorem\ref{Thm1}.\\
In particular 
\beq
u_0(t,y,p) = {\rm det}^{1/2}\Bigl(D-\Theta B)\Bigr)
\eeq
\end{theorem}
{\bf Sketch of proof.} 
These result can be proved following the same strategy as for proving Theorem \ref{PVVK}. \\
 We first prove the Theorem for some $\Theta$ ($\Theta=i\1$), following the proof of Theorem \ref{Thm1}.
   then we can get the Theorem for any $\Theta$
by the variation argument as in the proof of Theorem \ref{PVVK}. $L^2$ estimate for operator norm
of Fourier-Integral operators is used to control  the remainder terms. \hfill \sq
\begin{remark}
It is not difficult to adapt the proof of Theorem \ref{eht} concerning an Ehrenfest time estimate  to the setting of Theorem \ref{lasibu}.
\end{remark}

\end{document}